# Implication of Natal Care and Maternity Leave on Child Morbidity: Evidence from Ghana

Danny Turkson[1,2] & Joy Kafui Ahiabor[2]

[1] Economics Department, University of North Carolina at Greensboro, USA

[2] Economics Department, University of Ghana, Ghana

Correspondence: Danny Turkson, Bryan School of Business and Economics, University of North Carolina at Greensboro, USA. E-mail: d_turkso@uncg.edu

**Abstract**

The aim of government with the help of the Ghana Health Service (GHS) and other stakeholders has been to reduce the level of child morbidity which leads to child mortality in Ghana. This study on natal care and its implication on child morbidity would help the government in formulating appropriate policies to curb this problem. This study uses Acute Respiratory Infection (ARI) which is an infection of the lungs and respiratory tract as a proxy for child morbidity. The specific aim of this study is to ascertain the effect of Natal Care (Antenatal care, Delivery care and Post-natal care) and Maternity leave on Child Morbidity. The study employed data from the Ghana Demographic and Health Survey (2014) using the Probit estimation method to estimate the health, demographic and income factors that influence child morbidity in Ghana. It shows evidence that some stages of natal care, unpaid maternity leave, and other demographic factors have a significant impact on child morbidity in Ghana. Specifically, failure to receive post-natal care within first week of delivery causes a 3% increase in the possibility of ARI in children under five. The study also shows that a mother's income determines her health care purchases; in that an unpaid maternity leave causes a 3.9% increase in the possibility of ARI in children under five compared to a paid maternity leave.

## 1. Introduction

Stakeholders in the global health sector have made tremendous effort to enhance the health care of children all over the world and this has yielded some success. Although these successes are not at their all-time high as the reduction of child morbidity is still a daunting task to these experts. According to Kippenberg et al, (2005) about two-thirds of newborn or child deaths from the global perspective can be attributed to child infections, prematurity, and asphyxia which are illnesses that are preventable. 99% of these illnesses happen in less developed countries. Child morbidity is the prevailing cause of mortality among children though not all diseased children end up dying. The Sustainable Development Goals (SDGs) as part of the Agenda 2030 adopted by the General Assembly in 2015, seeks to reduce neonatal and under five years mortality to as low as 12 per 1000 live births and 25per 1000 births respectively to champion good health and wellbeing as a component of its seventeen goals. It also seeks to end the epidemic of tuberculosis and other diseases in all countries. To make this goal attainable, all countries need to put in more effort, especially low to middle-income countries where child mortality rates, particularly, newborn mortality rates are high.

Ghana Demographic and Health [GDHS] (2014) reported that pneumonia and other respiratory tract infections have been the major cause of child ill health and death in Ghana (Ghana Statistical Service [GSS], 2014). ARI has a wide range of effects like bacterial and viral infection of the lungs and respiratory tracts. Other triggers include exposure to air pollution, low birth weight, malnutrition, and overcrowding, which are all important risk factors. Acute Respiratory Infections (ARI) contribute to almost 33 percent of all deaths in pre-school children in developing countries. Exposure to outdoor concentrations of air pollution has been pointed out as one of the possible risk factors associated with respiratory diseases (Gouveia & Fletcher, 2000). Mother's report from the GDHS (2014) estimated 4 percent ARI symptoms in children under the age of five, but it is necessary to note that children in remote areas are likely to experience symptoms of ARI two times more than their counterparts living in urban areas.

Several studies have outlined problems confronting child health with relation to post-natal care of children including neonates and infants as emanating from diseases such as Diarrhea, Pneumonia, Asphyxia, Malaria among many others. Some studies have cited that more than one million child mortalities as a result of morbidity that has occurred in Africa are within the first week of life, half of them on the first day of life. Fengxiu Ouyang et al. (2013) claimed that between 3 to 4 million stillbirths occur annually across the world as a result of maternal morbidity and almost 97-99% are found in less developed countries. The continuous incidents and cases of childhood diseases and mortality in Ghana have generated a lot of concern among various stakeholders of public health in the country.

The perinatal rate in Ghana as at 2014 was 38/1000 of all pregnancies in the five years prior to the survey (GSS, 2014). Prior to the Ghana Demographic and Health Survey (GDHS, 2014), they identified about 4% of pre-school children with ARI symptoms. The issue of concern is the continuous morbidity of children in the country through Acute Respiratory Illnesses such as pneumonia bronchiolitis in the phase of interventions in post-natal care for children. Perhaps, it is the emphasis on the antenatal and delivery care for the children or an increase in a mother's income level to purchase proper health care that could offer a drastic reduction in the morbidity of children in the country.

Dealing with child morbidity also requires that we deal with the health of the mothers as well since the health of the mother will have a direct influence on the health status of an unborn child during the antenatal and the delivery stages specifically. The labor supply of the mother hence her income level affects child morbidity. Research has shown that improvement in the prenatal care and nutrition levels of the mother has a direct influence on maternal labor productivity especially among vulnerable families (Tambi, 2017). If there is the provision of quality clinical services to the mother during this stage, it can go a long way to impact on her labor supply and income levels and the health of the child (Mahiti et al., 2015)

Studies have examined that maternal employment affects the health status of low-income, elementary-school-aged children using instrumental variables estimation and experimental data from a welfare-to-work program implemented in the early 1990s. In a secondary analysis using fixed effects techniques on longitudinal survey data collected in 1998 and 2001, we find a comparable adverse effect of maternal employment on child health that supports the external validity of our primary result. It was found that employment among low-income women have a modest adverse effect on the general health status of young children (Gennetian et al., 2010). It is based on this reason the study sought to explore the impact of natal care encompassing antenatal, delivery and post-natal care and a maternity leave on the morbidity of children.

*1.1 Research Objective*

The main aim of this research is to explore the impact of natal care (antenatal, delivery and post-natal care) and maternity leave on child morbidity. In achieving this, it is prudent to identify the causes of child morbidity in Ghana. This study is centered on children aged 0 to 5 years because, it is the vulnerable age range for child morbidity in Ghana (UNICEF, 2012). The scope of the study is however broad as it is measured by the number of visits to the antenatal clinic and post-natal clinic before and after delivery respectively in Ghana. This is specifically centered on Acute Respiratory Illness of the children as the proxy for measuring child morbidity.

## 2. Literature Review

The child morbidity rate may be defined as the number of sick children that occur per thousand live births in any population in one calendar year. From this definition, it is obvious that the child morbidity rate does not consider stillbirth, but only live births and infants (UN, 2015). Child morbidity rate comprises of two parts viz. neo-natal morbidity rate and postneonatal morbidity rate. The neonatal morbidity rate also comprises of two parts viz. early neonatal morbidity rate and late neonatal morbidity rate. Neonatal morbidity is the likelihood of falling sick in the first month (28 days) after birth and post-neonatal mortality is the likelihood of falling sick after the first month of birth but before the 12$^{th}$ month. Therefore, focusing on infant morbidity as a single measure disguises the unequal share of neonatal and post-neonatal morbidity. According to one estimate, 50-60 percent of infant deaths i.e. nearly 5 million neonates die each year in the world as a result of morbidity, of which 96 percent are in developing countries (UN, 2015).

Natal Care for children has been classified under various types namely prenatal care, delivery care and postnatal care. Stout (1997) defined natal care as any health care services provided to a pregnant woman until birth. This definition is based on her trace of the historical precedence and aim of natal care. Although she clarified it was a response to detect and prevent eclampsia, it is too basic and may fail to capture the broad perspectives of care for which natal care presently undertakes. A more appropriate definition that portrays the aims of current natal care was given by Klerman (1990). He acknowledged that natal care is not a single intervention but an integration of several factors composing of early and continuing risk assessment, health promotion, and medical and psychosocial interventions and follow-ups. This definition absorbs better the essence and objectives of natal care and coincides with that given by World Health Organization (WHO). According to WHO, the reason for natal care is to screen predominantly healthy women to detect early signs of or risk factors for abnormal conditions or diseases and to follow this detection with effective and timely intervention (Carroli, Rooney, & Villar, 2001; Lumbiganon et al., 2004).

In summing up, natal care can be defined as a holistic, systematic, and evidence-based approach to care during pregnancy, targeted at ensuring the well-being of the woman and fetus, enhance women's positive posture towards childbirth, and prepare a safe and good starting point for the newborn child (Dragonas'et al., 1998; WHO, 2003). It must be timely focused, contextually sensitive, and individually specific, following general recommendations. It covers all medical, physical, and socio-demographic examinations and risk assessments during pregnancy. Health education and counseling services have also been included in a parental care package so that attitudinal, behavioral and sociodemographic risks surrounding pregnancy, childbirth, and the postpartum period can be managed. This is why not any care provided during pregnancy can be classified as health care but targeted care, administered by well-trained personnel, taking into consideration the various factors that may affect the process of pregnancy and childbirth. It is anticipated that when the various components of prenatal care are well coordinated, there would be positive effects, not only in pregnancy and labor but on the behavioral patterns of women towards maternal health care and care for the newborn.

*2.1 Empirical Literature on Factors Influencing Child Morbidity*

Some demographics has been identified to affect child morbidity. A study conducted by Ntimba and Mbago (2005) showed that the age of a mother can significantly explain the occurrence of infant and child mortality. Mothers who had their first child before the age of twenty are 2.4 times more likely to have child death than those who had their first child between the ages 20-34. Generally young mother that are below twenty years and those between ages 40-49 had a higher chance of child mortality than those born to mothers aged 20-39 years (Mustafa & Odimegwu, 2008: Kembo & Ginneken, 2009). Dede (2013) indicated that, among the children who died at the age of 0-11months, mothers who were aged between 25-29 years had more child deaths than mothers who were aged between45-49 years contrary to other studies. This result does not conform to other studies where women aged less than 20 years and 40+ years are expected to experience a high rate of IMR (Schneider et al., 2012).

Over the years, mortality levels in rural areas are consistently higher than those in urban areas. Some existing literature have found that mothers living in rural areas have a relatively higher chance of infant mortality than mother's living in urban areas (Bello & Joseph, 2014). Another paper by Mostafa (2012) found that place of residence has a significant association with neonatal mortality and child survival, where people in rural areas die more than people in urban areas. Results from a research by Dede (2013) on the determinants of infant mortality in Tanzania shows that place of residence has no significant relationship with infant mortality. The study further states that 5.6% of infant mortality are from residence in urban areas while 4.5% deaths are from residents in rural areas which contradicts Mostafa's earlier accretion. Also, a study by Mostafa (2012) showed that the education of a mother influences her choice and skills in health care practices and is more prudent in ensuring neonatal mortality. Educated mothers are likely to have more antenatal visits as to uneducated mother (Mullany, 2007). A paper by Nketiah-Amponsah, Boakye-Yiadom and Agyemang (2016) shows that higher maternal education significantly and consistently reduces the incidence of diarrhea, ARI, and fever among children aged under-five in Ghana. Contrary to the above studies, Dede (2013) showed results that infant mortality is more frequent among mothers with complete and higher educational attainment as against mothers with incomplete primary educational attainment. This result may be attributed to the fact that mothers with secondary or higher education have little time to take care of their newborn babies.

Economic factors do affect the choice of medical care and hence influences child morbidity. A study by Arthur (2012) using the GDHS (2008) showed that in Ghana wealth has a significant influence on the adequate use of antenatal health care. Shrestha (2012) indicated that about 50% of poorest women had no Antenatal Care (ANC) in Nepal. Similarly, Sharma (2002) showed that the percentage of women with adequate visits for ANC (minimum 4 times) increased from 4 percent (low economic status) to 42 percent among high economic status women in Nepal. Wealth status also affects child mortality rates in Ghana. Children in households of the highest wealth quintile have the lowest mortality rates for both child mortality and under-five mortality (Agoe, 2013). Dede's (2013) study on the determinants of infant mortality in Tanzania showed results that 5.2% of the deaths occurred to children born by mothers who were employed while 3.8% occurred to children born by mothers who were not employed in the period of one year preceding the survey in 2010. This brings to bear the importance of maternity leave, to enable mothers take time off to cater for their children. The discussion now is if these mothers should be entitled to a salary during maternity leave. Heymann et.al (2011) study revealed that paid maternity leave significantly reduces child mortality. The results also showed that family income is a strong predictor of infant and child morbidity and mortality rates. It is thereby convenient to use paid maternity leave as a proxy for the income level of a mother in relation to child morbidity.

A study by Hollm-Delgado et al. (2014) to determine whether Bacille Calmette Guerin (BCG) vaccination is linked to the risk of ARI among children under 5 years of age showed results that BCG vaccination was associated with a 17% to 37% risk reduction of ARI and hence concluded that BCG had a significantly lower risk of suspected ARI. Antenatal care (ANC) is the care that a woman receives throughout her pregnancy and some weeks post-partum. The reason for providing ANC is to screen mostly healthy pregnancies to detect early signs and risk factors for abdominal conditions or diseases and also to follow this detection with effective and timely intervention (Lumbiganon et al., 2004). Antenatal care has the potential of solving ill health issues, yet a lot of mothers find it difficult to access it. A theory by Phillippi and Roman (2013) explains this situation is the motivation and facilitation theory of prenatal care. Post-natal period is crucial to a mother and child's health and survival. The hours and days after the delivery of a child is seen as the most vulnerable time of a child. The lack of proper health care can lead to death or disability in infants and children. 25 percent of child deaths occur in the first month after delivery. These deaths usually occur before childcare health assistants start to provide care which is usually in the sixth week when immunization takes place. Low health care attention during the post-natal stage has an unwanted influence on maternal, newborn and child health (MNCH) programmes along the continuum of care. A study using Indonesia Demographic Health Survey to determine the risk of early neonatal mortality used deaths occurring within the first 7 days of life as outcome variables for the study and showed results that neonatal death out of the entire under-five child deaths recorded increased from 66% in 1994 to 78% in 2007 (Titaley, Dibley, & Roberts, 2011).

## 3. Methodology

### 3.1 Conceptual Framework

Deaton et al. (2005) conceptualize the determinants of child morbidity in three main factors. These are demographic, health care systems and economic factors. For this research, we would consider some health factors which include Antenatal care, Delivery care, Postnatal care, vaccination, death of children and birth experience as estimates in determining the form of health care given to mother and their child before, during and after childbirth. Other demographic factors are necessary inclusions in this estimation. These factors include mother's age, child's age, wealth, location, mother's education, and marital status. Also concerning economic factors, the study would use the income of the mothers as a measure.

$$child\ morbidity = f(health\ factors, economic\ factors, demographics)$$

### 3.2 Estimation Model Technique

The dependent variable is the key determinate in the choice of model to use. In cases where the dependent variable is binary, the goal would be to find the probability of something happening. Therefore, the qualitative response regression models that emerge from the normal cumulative distribution function (CDF) is what is known as the Probit model which is sometimes called the normit model (Gujarati, 2004). For this study, Child Morbidity either occurs or does not occur, which is a qualitative response. The binary regression model is often used for these types of study. Resulting from Balangun and Yusuf (2011), the binary probit model through the maximum likelihood estimation method suits this study. The probit model is preferred to the linear probability model (LPM) because the LPM does not constrain the probabilities to lie within the range of 0 and 1.

The study then specified the model as:

$$Pr(Child\ Morbidity = 1|X_i) = \phi(\beta_k X_i + \varepsilon_i)$$

Where

- **Child Morbidity** represents the probability that a child suffers an Acute Respiratory Infection (ARI).
- $\phi$ represents the cumulative standard normal distribution function
- $\beta_k$ is a vector of parameters.
- $X$ is vector of independent variables {*Mother's Age, Location, Wealth, Child's Age, Birth Experience, Antenatal care, Marital Status, Mother's Education, Dead Children, Place of Delivery, Post-natal Care,*

*Vaccination, Maternity Leave*}

- $\varepsilon_i$ is the error term

### 3.3 Definition of Variables

*Child Morbidity*: This dependent variable child morbidity, that is, acute respiratory illness is presented as a dummy and has the value of '1' when the child has acute respiratory illness and '0' when the child does not have acute respiratory disease.

*Mother's Age*: This variable shows how old a mother is. It is a continuous variable which is measured in years and has an expected sign of either negative or positive.

*Child's Age*: This variable as the name denotes is the age of the child.

*Location:* This variable shows the location of the household which is either urban or rural. It is expected that a household in an urban area is less likely to have their child experiencing ill health since good health care is readily available to access.

*Marital Status*: This variable tells us the marital status of the mother to determine if she has any form of assistance from a spouse in taking care of the child.

*Mother's Education*: This variable centers on a mother's level of education, that is, if the mother is educated (at least a primary education) or uneducated. The higher a mother's level of education the greater the exposure she has to child health care practices and the less likely it is for the child to experience ill health.

*Wealth:* This variable measures the financial worth of the family. It is a categorical variable; Poorest, Poorer, Middle income, Richer and Richest. We expect that the richer a family is the easier it is to afford good health care and the less likely it is for the child to be of ill health.

*Birth Experience*: This is the number of successful pregnancies. This suggests that the more birth experiences a mother has the less likely it is for her child to be of ill health.

*Dead Children*: This variable is the number of child death between the age of zero and five years in a household.

*Antenatal Care*: This variable measured as the number of antenatal care visits made during pregnancy. As a mother goes for all her required antenatal appointments the less likely it would be for that baby to be sick when delivered.

*Vaccination*: It measures if the child has received BCG vaccine and not; which means a child who has received the BCG vaccine is less likely to experience ill health in the first five years.

*Place of Delivery*: This variable demonstrates the nature and environment in which child delivery occurs. The variable is presented as a modern health facility (public, private and other health facilities) or a traditional environment (home delivery and others). Children born through modern means are less likely to experience ill health during the first five years of their life.

*Postnatal Care*: The first 7 days of an infant's life is very crucial in determining a child's health (Titaley, Dibley & Roberts, 2011). This variable is preferably presented as first postnatal visit took place after the first week of delivery or within the first week of delivery. The longer the first postnatal health care visit the more likely it is for a child to have ill health.

*Maternity Leave*: This variable is to represent the financial state of the mother within the time of natal care. Paid maternity leave increases the income available to both parents and children during a critical time and secures a mother's employment after childbirth to protect from suffering long-term wage penalty on individual income and household income (Heymann, Raub & Earle, 2011). For women in the labor force it is necessary to take some time off during the period of childbirth to cater for child breastfeeding and immunization (Heymann, Raub & Earle, 2011). This variable seeks to determine the effect of paid and unpaid labor income or wages on the health of the child.

### 3.4 Source of Data

Data was obtained from the 2014 Ghana Demographic Health Survey (GDHS). The survey is a nationally represented survey cutting across all the ten regions in Ghana populated by 9396 women between the ages of 15-49 and 4388 men between the ages 15- 59 from interviewed households amounting to 11 835. The objective of the survey was to provide recent and reliable information on housing characteristics, child mortality, maternal and child health, nutrition and mothers labor income. The survey provides information among children aged between 0-59 months in relation to child disease treatment, prevention and prevalence, and adult disease as well. The survey followed a two-stage sample design. Stage one involved selecting sample points (clusters), that is 427 clusters in total consisting of 216 in urban and 211 in rural areas. Stage two involved the household systematic sampling. A household listing operation was undertaken in January-March 2014, and households to be included in the survey were randomly selected from the list (GSS, GHS & ICF International, 2015).

## 4. Results and Discussions

From the survey, the sample size of 3057 responded. 2314 respondents, which forms 76% of the entire sample size recorded no cases of respiratory illness whereas 743 which is 24% recorded cases of respiratory diseases. The descriptive statistics of the independent variables is presented in Table 1 below. The level of Antenatal care provided is determined in this research by the number of antenatal visits respondents made during their time of pregnancy. Respondents in the sample size made about 7 antenatal care visits before the child was born. Place of delivery determines the level of delivery care, that is, either a modern health facility or at home through traditional means. The survey showed that about 73% of respondent had their children in health facilities and the remaining 27% through traditional means. The nature of post-natal care is determined by the time of the first visit after delivery. About 63% of the sample population had their first post-natal visit after the first week of delivery and the remaining 37% in the first week after delivery. For children with ARI, about 65% of them are children who had their post-natal checkup after a week of delivery. Maternity leave is the period a mother takes off work during pregnancy and after the childbirth (Heymann et.al, 2011). The employer either chooses to pay or not pay these mothers during their leave period. Result for this study has shown that about 5.8% of mothers out of the entire sample were entitled to a paid leave and 12.2% of mothers were not entitled to a paid leave, whilst the rest are not entitled to a leave.

Table 1. Descriptive statistics of independent variables

| Variables | Total Sample | | Child Morbidity Cases | |
|---|---|---|---|---|
| | Frequency (N=3057) | Percentage (%) | Frequency (n=743) | Percentage (%) |
| **Location** | | | | |
| • Urban | 1307 | 42.7 | 314 | 42.3 |
| • Rural | 1750 | 57.3 | 429 | 57.7 |
| **Marital Status** | | | | |
| • Married | 2617 | 85.6 | 629 | 84.7 |
| • Not Married | 440 | 14.4 | 114 | 15.3 |
| **Mother's Education** | | | | |
| • Educated | 2032 | 66.5 | 471 | 63.4 |
| • Uneducated | 1025 | 33.5 | 272 | 36.6 |
| **Wealth** | | | | |
| • Middle Income & Above | 1444 | 47.2 | 328 | 44.2 |
| • Below Middle Income | 1613 | 52.8 | 415 | 55.8 |
| **Place of Delivery** | | | | |
| • Modern | 2236 | 73.1 | 522 | 70.3 |
| • Traditional | 821 | 26.9 | 221 | 29.7 |
| **Post Natal Care** | | | | |
| • 1 week and beyond | 1924 | 62.9 | 485 | 65.3 |
| • Less than a week | 133 | 37.1 | 258 | 34.7 |
| **Vaccination** | | | | |
| • Yes | 2472 | 80.9 | 601 | 80.9 |
| • No | 585 | 19.1 | 142 | 19.1 |
| **Maternity Leave** | | | | |
| • No | 2508 | 82.0 | 589 | 79.3 |
| • Paid Leave | 177 | 5.8 | 43 | 5.8 |
| • Unpaid Leave | 372 | 12.2 | 111 | 14.9 |

|  | Mean | Standard Dev. | Mean | Standard Dev. |
|---|---|---|---|---|
| **Mother's Age** | 30.75 | 7.03 | 30.90 | 7.11 |
| **Child's Age** | 1.53 | 1.31 | 1.64 | 1.25 |
| **Birth Experience** | 3.38 | 2.10 | 3.41 | 2.12 |
| **Dead Children** | 0.26 | 0.60 | 0.31 | 0.64 |
| **Antenatal care** | 6.80 | 6.53 | 6.64 | 4.33 |

*Source*: Author's compilation based on GDHS 2014.

*4.2 Probit Regression Estimates of the Determinants of Child Morbidity*

From Table 2, the marginal effect shows that an increase in the age of a child by one year will increase the child's probability of contracting an acute respiratory illness by 1.4%. One possible reason for this finding may be attributed to the fact that as children grow into toddlers and age further, they begin to play most of the time and easily pick up germs and bacteria. The surging of environmental pollution in Ghana, poses a higher risk of contracting respiratory infections in children as their immune systems are still developing. Also, an increase in the infant children deaths of a household increases the probability of child morbidity in that household by 3.9%. It is logical to assume that the cause of death other than by accident is illness. For a household experiencing an increasing number of child deaths, there is a high possibility of these deaths being attributed to a familiar illness among these children. This means living children of that household may also have this illness as well. The chances of an accident reoccurring among children in the same household are likely to be marginally low as compared to diseases.

The first post-natal visit of a child beyond seven days after delivery increases the probability of that child having an acute respiratory infection by 2.9%. This shows that the first few days after delivery of an infant life is very crucial and for this reason, it is necessary that a mother ensures the child gets health care check from some medical personnel at least within the first week after delivery to determine and certify the child's wellbeing. The longer it takes for a child to have his/her first post-natal care, the higher the likelihood of him contracting an illness. Furthermore, mothers who go on maternity leave without pay have the likelihood of their children having a case of acute respiratory cases as compared to those who do not go for leave. The marginal effect shows that for every unpaid maternity leave of a mother, the probability that the child would suffer from child illness is 3.9% as compared to mothers with paid maternity leave. Access to good antenatal health care during pregnancy requires high monetary expenses, and this explains why the richer and richest households have lower child morbidity rate.

Table 2. Probit Model for Determinants of Child Morbidity in Ghana

Dependent Variable: Child Morbidity

| Variables | Coefficient | Marginal Effect | P-value |
|---|---|---|---|
| Mother's Age | 0.0057 | 0.0017 | 0.845 |
| Child's Age | 0.0441 | 0.0136** | 0.027 |
| Location |  |  |  |
| • Urban | 0.1039 | 0.0206 | 0.118 |
| Marital Status |  |  |  |
| • Married | -0.0486 | -0.0150 | 0.516 |
| Mother's Education |  |  |  |
| • Educated | -0.0518 | -0.0160 | 0.407 |
| Wealth |  |  |  |
| • Poorer | -0.0291 | -0.0094 | 0.690 |
| • Middle | -0.0640 | -0.0204 | 0.440 |
| • Richer | -0.2014 | -0.0613** | 0.035 |
| • Richest | -0.1906 | -0.0582* | 0.076 |
| *Birth Experience* | -0.0336 | -0.0103 | 0.122 |
| *Dead Children* | 0.1280 | 0.0396*** | 0.008 |
| *Antenatal care* | -0.0021 | -0.0006 | 0.547 |
| *Vaccination* | 0.0273 | 0.0085 | 0.674 |
| *Place of Delivery* | | | |
| • Modern | -0.0717 | -0.0222 | 0.244 |
| *Post Natal Care* | | | |
| • 1 week and beyond | 0.0954 | 0.0296* | 0.067 |
| *Maternity Leave* | | | |
| • Paid Leave | 0.0794 | 0.0248 | 0.488 |
| • Unpaid Leave | 0.1959 | 0.0635** | 0.011 |
| *Con* | -0.7685 | | |

Number of Obs = 3,057; Wald chi2 (18) = 35.50; Prob>chi 2= 0.0082; Pseudo R2= 0.0106; Log likelihood = -167704288.

*Note*. ***, **, * significant at 1%, 5% and 10% respectively.

### 4.3 Natal Care and its influence on Child Morbidity

The objective of this study is to define natal care and its implications on child morbidity by focusing individually on the three stages of natal care - Antenatal care, Delivery Care and Post-natal care - to identify their individual effects on child illnesses. Results have shown that of the three natal care variables (antenatal care, delivery care, and post-natal care) explored, only post care has a significant effect on child morbidity (Acute Respiratory Infection). The three stages of natal care have two stages dependent on the health state of the mother being trickled down to the child's health and the remaining one being directly the health state of the child. Considering the fact that respiratory illnesses are caused by viruses and bacteria that affects the lungs and respiratory tracts which are contracted when exposed to polluted environment, it is logical to conclude that the impact of Antenatal and delivery care on a child's health is dependent on the health status of the mother and does not have direct effect on a child; this may be the reason why the results for these two variables were not significant. Post-natal care on the other hand which is the childcare after delivery from Table 2 was significant at 10% and positively related to the child morbidity. When children are born their immune system is gradually building up and for this reason, any exposure to a polluted environment may result in bacterial infection in their lung and respiratory tracts. This is why it is important to have a postnatal checkup within a week after delivery and regularly to quickly identify any form of infection. Children born to mothers who had a postnatal checkup a week or more after delivery are approximately 2 percentage points more likely to experience ARI. This result confirms a study conducted by Titaley, Dibley & Roberts (2011) which used Indonesia Demographic Health Survey to determine the risk of early neonatal mortality. They used deaths occurring within the first 7 days of life as outcome variables, to determine how neonatal death increased from 66% in 1994 to 78% in 2007 out of all under-five child death recorded.

### 4.4 Mother's Income during Maternity Leave and its implication on Child Morbidity

The health care service one may acquire is highly dependent on the income of the respondent during and after pregnancy (Heymann, Raub & Earle, 2011). Just like the wealth of a household, the household health care purchase decision is dependent on the wealth or income level of respondents. To determine the impact of natal care on child illnesses, it is necessary to consider the income level of the mother during pregnancy. For a working mother who goes on a maternity leave without pay, the results show that the coefficient of unpaid maternity leave is significant and positively related to child illnesses. This means that children born to mothers on unpaid maternity leave are 6 percentage points more likely to experience ARI. This is because an unpaid mother on maternity leave suffers from a reduction in household income and hence cannot afford to purchase the best health care services before, during and after childbirth (Heymann, Raub, & Earle, 2011). This affects her frequency of antenatal care, the place of delivery and also the time of first postnatal care visit.

## 5. Conclusion and Recommendation

### 5.1 Conclusions

The study revealed that for the three stages of natal care; post-natal care does have an effect on child health, and not antenatal care and delivery care. This is because for the first and the second stages of natal care, the health of the child is predominately dependent on the health of the mother and for this reason, proper care is taken by the mother to ensure that she and her child are in good health. The mother, in this case, has full control in ensuring that she and her baby are healthy before delivery. The baby is not directly exposed to the environment because the mother shields the child from any harm, but after delivery, the baby is directly exposed to environmental hazards like bacterial and viral infections that affect the baby directly. Hence, without proper post-natal care, the child would be exposed to germs which may get into his/her lungs and respiratory tract and cause respiratory infection of the child. The longer it takes for a child to have his/her first post-natal visit, the higher the child stands at the risk of being ill by 2.9%.

This study revealed that an increase in the number of child deaths a household experiences increases the probability of a child in that household being ill by 3.9% point. The study revealed that a child born to a mother who has experienced repeated death of children has a high likelihood of being ill. This shows that there is a significant positive relationship between child mortality and child morbidity as mentioned in the earlier chapters of this study. Some demographic factors are significant and hence affect child morbidity. One of these factors is the child's age, which has a positive relationship with child respiratory illness. An increase in the age of a child increases the probability of the child having a respiratory illness by 1.4% points. As the child grows, he/she is more exposed to our polluted environment and therefore is exposed more to bacterial infections which may affect the lungs and respiratory tract that causes child respiratory illnesses. Unpaid maternity leave is positively related to child morbidity. This finding is supported by the fact that to access the best natal health care demands a lot on financial resources and it is difficult for a mother with an unpaid maternity leave to make the best health care purchases needed.

*5.2 Recommendations*

A couple of observations emerged from the study which requires close attention by government and other stakeholders in reducing the level of child morbidity. Post-natal care in Ghana was found to be positive and significant to the occurrence of ill health of children in Ghana. Mothers and health professionals should focus more on the provision of postnatal care from skilled health professionals and also ensure timely health checkup after delivery. Government and its stakeholders can help facilitate this by increasing the education and awareness of the need for a timely post-natal checkup to mothers and health professionals. Since unpaid leave increases the probability of ill health, policy recommendation is to enforce paid maternity leave laws in institutions to protect mothers. This would help mothers assess quality natal care without financial constraints.

**Acknowledgments**

Special thanks to Gloria Naana Nissi Diafo for critiquing and proofreading the manuscript.

**Competing Interests Statement**

The authors declare that there are no competing or potential conflicts of interest.

**References**

Agoe, P. O. (2013). Attaining the Millennium Development Goal 4 of Reducing Child Mortality in Ghana: The Role of Foreign Aid in Attaining MDG 4 in Ghana.

Arthur, E. (2012). Wealth and Antenatal Care Use: Implications for Maternal Health Care Utilisation in Ghana. *Health Economics Review, 2012*. https://doi.org/10.1186/2191-1991-2-14

Balogun, O. L., & Yusuf, S. A. (2011). Determinants of Demand for Microcredit Among the Rural Households in South-Western States, Nigeria. *Journal of Agricultural and Social Science, 7*, 41-48.

Bello, R. A., & Joseph, A. I. (2014). Determinants of Child Mortality in Oyo State, Nigeria. *African Research Review, 8*(1), 252-272. https://doi.org/10.4314/afrrev.v8i1.17

Carroli, G., Rooney, C., & Villar, J. (2001). How effective is antenatal care in preventing maternal mortality and serious morbidity? An overview of the evidence. *Paediatric and Perinatal Epidemiology, 15*(Supplement 1), 1-42. https://doi.org/10.1046/j.1365-3016.2001.0150s1001.x

Dede, K. S. (2013). *Determinants of infant mortality in Tanzania* (Thesis (MA)-University of Ghana, 2013).

Ouyang, F., Zhang, J., Betrán, A. P., Yang, Z., Souza, J. P., & Merialdi, M. (2013). Recurrence of adverse perinatal outcomes in developing countries. *Bulletin of the World Health Organization*, *91*, 357-367. https://doi.org/10.2471/BLT.12.111021

Gennetian, L. A., Hill, H. D., London, A. S., & Lopoo, L. M. (2010). Maternal Employment and The Health of Low-Income Young Children. *Journal of Health Economics, 29*(3), 353-363. https://doi.org/10.1016/j.jhealeco.2010.02.007

Ghana Statistical Service (GSS), Ghana Health Service (GHS), and ICF International. (2015). *Ghana Demographic and Health Survey 2014*. Rockville, Maryland, USA: GSS, GHS, and ICF International.

Ghana Statistical Service [GSS]. (2008). Ghana Health Services (GHS), ICF Macro: *Ghana Demographic and Health Survey (2008) Report. ICF Macro*.

Ghana Statistical Service [GSS]. (2014). *Ghana Demographic and Health Survey - Final Report-DHS Program, ICF International*, Rockville, Maryland, USA. 2015.

Mahiti, G. R., Mkoka, D. A., Kiwara, A. D., Mbekenga, C. K., Hurtig, A. K., & Goicolea, I. (2015). Women's perceptions of antenatal, delivery, and postpartum services in rural Tanzania. *Global health action*, *8*(1), 28567. https://doi.org/10.3402/gha.v8.28567

Gouveia, N., & Fletcher, T. (2000). Respiratory diseases in children and outdoor air pollution in Sao Paulo, Brazil: a time series analysis. *Occupational and environmental medicine*, *57*(7), 477-483. https://doi.org/10.1136/oem.57.7.477

Gujarati, D. N. (2005). *Basic Econometrics* (4th ed.). McGraw-Hill Book Company. New York

Heymann, J., Raub, A., & Earle, A. (2011). Creating and using new data sources to analyze the relationship between social policy and global health: the case of maternal leave. *Public Health Reports*, *126*(3_suppl), 127-134. https://doi.org/10.1177/00333549111260S317

Hollm-Delgado, M. G., Stuart, E. A., & Black, R. E. (2014). Acute lower respiratory infection among Bacille Calmette-Guérin (BCG)–vaccinated children. *Pediatrics*, *133*(1), e73-e81. https://doi.org/10.1542/peds.2013-2218

Kembo, J., & Van Ginneken, J. K. (2009). Determinants of Infant and Child Mortality in Zimbabwe: Results of Multivariate Hazard Analysis. *Demographic Research, 21*, 367-384. https://doi.org/10.4054/DemRes.2009.21.13

Knippenberg, R., Lawn, J. E., Darmstadt, G. L., Begkoyian, G., Fogstad, H., Walelign, N., ... & Lancet Neonatal Survival Steering Team. (2005). Systematic scaling up of neonatal care in countries. *The Lancet*, *365*(9464), 1087-1098. https://doi.org/10.1016/S0140-6736(05)74233-1

Klerman, L. V. (1990). The Need for a New Perspective on Prenatal Care. In: I. R. Merkatz, & J. E.Thompson, (eds.), *New Perspectives on Prenatal Care*. New York: Elsevier 6.

Lumbiganon, P., Winiyakul, N., Chongsomchai, C., & Chaisiri, K. (2004). From Research to Practice: The Example of Antenatal Care in Thailand. *Bulletin of the World Health Organization, 82*(10), 746-749.

Mostafa, K. S. M. (2012). Maternal Education as a Determinant of Neonatal Mortality in Bangladesh. *Journal of Health Management, 14*(3), 269-281. https://doi.org/10.1177/0972063412457509

Mullany, B. C., Becker, S., & Hindin, M. (2007). The Impact of Including Husbands in Antenatal Health Education Services on Maternal Health Practices in Urban Nepal: Results from a Randomized Controlled Trial. *Health Education Rresearch, 22*(2), 166-176. https://doi.org/10.1093/her/cyl060

Mustafa, H. E., & Odimegwu, C. (2008). Socioeconomic Determinants of Infant Mortality in Kenya: Analysis of Kenya DHS 2003. *Journal of Humanities & Social Sciences, 2*(2), 1-16.

Nketiah-Amponsah, E., Boakye-Yiadom, L., & Agyemang, M. (2016). The effect of maternal education on child health: some evidence from Ghana. *International Journal of Economics and Business Research*, *11*(4), 366-385. https://doi.org/10.1504/IJEBR.2016.077027

Ntimba, G. L., & Mbago, M. C. (2005). *Some Socio-Economic and Demographic Determinants of Infant and Child Mortality in Tanzania: A Case Study of Karagwe District, Kagera Region*.

Phillippi, J. C., & Roman, M. W. (2013). The Motivation-Facilitation Theory of Prenatal Care Access. *Journal of Midwifery & Women's Health, 58*(5), 509-515. https://doi.org/10.1111/jmwh.12041

Ruhm, C. J. (2000). Parental Leave and Child Health. *Journal of Health Economics, 19*(6), 931-960. https://doi.org/10.1016/S0167-6296(00)00047-3

Schneider, J., Kondareddy, D., Gandham, S., & Dude, A. (2012). Using Digital Communication Technology Fails to Improve Longitudinal Evaluation of An HIV Prevention Program Aimed at Indian Truck Drivers and

Cleaners. *AIDS and Behavior, 16*(5), 1359-1363. https://doi.org/10.1007/s10461-011-0060-6

Sharma, M. B. R. (2002). *Factors Affecting Utilization of Antenatal Care Services in Nepal*. Mahidol University

Shrestha, G. (2012). Statistical Analysis of Factors Affecting Utilization of Antenatal Care in Nepal. *Nepal Journal of Science and Technology, 12*, 268-275 https://doi.org/10.3126/njst.v12i0.6512

Stout, A. E. (1997). Prenatal Care for Low-Income Women and the Health Belief Model: a New Beginning. *Journal of Community Health Nursing, 14*(3), 169-180. https://doi.org/10.1207/s15327655jchn1403_4

Tambi, M. D. (2017). Children's Health, Maternal Labour Supply and Wealth Accumulation: Theory, Evidence and Policy Approach. *Health Econ Outcome Res Open Access, 3*, 135. https://doi.org/10.4172/2471-268X.1000135

Titaley, C. R., Dibley, M. J., & Roberts, C. L., (2012). Type of Delivery Attendant, Place of Delivery and Risk of Early Neonatal Mortality: Analyses of the 1994–2007 Indonesia Demographic and Health Surveys. *Health Policy Plan, 27*(5), 405-16. https://doi.org/10.1093/heapol/czr053

UNICEF. (2012). The State of The World's Children 2012: Children in an Urban World. *Social Sciences*.

United Nations. (2015). Child Mortality: Millenium Development Goal (MDG) 4. *WHO Report.* Retrieved from: https://www.who.int/pmnch/media/press_materials/fs/fs_mdg4_childmortality/en/